\documentclass[3p, times,merge]{elsarticle}

\usepackage[plain]{algorithm}
\usepackage{amsmath}
\usepackage{amssymb}
\usepackage{bm}
\usepackage[nolabel]{showlabels}
\usepackage[ISO=false]{diffcoeff}
\usepackage{listings}
\usepackage{graphicx}
\usepackage[utf8]{inputenc}
\usepackage{tablefootnote}
\usepackage{threeparttable}
\usepackage{makecell}
\usepackage{xcolor}
\usepackage{textcomp}
\definecolor{mydarkgray}{gray}{0.2}
\definecolor{darkgreen}{rgb}{0,.5,.0}

\usepackage{geometry}
\usepackage{subcaption}

\newcommand{\avg}[1]{\left\langle{#1}\right\rangle}
\newcommand{\qt}{{q_{\theta}}}

\lstdefinelanguage{MyPython}{
    keywords={def, for, int, float, return, if, else, range, product},   
    keywordstyle=\color{red}\bfseries,
    comment=[l]{\#},             
    string=[b]"                 
}

\begin{document}

\begin{frontmatter}

\title{Hierarchical autoregressive neural networks in three-dimensional statistical system}
\author[iis]{Piotr Białas}
\ead{piotr.bialas@uj.edu.pl}
\author[msip,ds]{Vaibhav Chahar}
\ead{vaibhav.chahar@doctoral.uj.edu.pl}
\author[ift]{Piotr Korcyl}
\ead{piotr.korcyl@uj.edu.pl}
\author[ift]{Tomasz Stebel}
\ead{tomasz.stebel@uj.edu.pl}
\author[ift]{Mateusz Winiarski}
\ead{mateusz.m.winiarski@student.uj.edu.pl}
\author[iis,ds]{Dawid Zapolski}
\ead{dawid.zapolski@doctoral.uj.edu.pl}

\address[iis]{Institute of Applied Computer Science, Jagiellonian University, ul.~Łojasiewicza 11, 30-348 Kraków, Poland}

\address[msip]{M. Smoluchowski Institute of Physics, Jagiellonian University, ul.~Łojasiewicza 11, 30-348 Kraków, Poland}

\address[ds]{Doctoral School of Exact and Natural Sciences, Jagiellonian University, ul.~Łojasiewicza 11, 30-348 Kraków, Poland}

\address[ift]{Institute of Theoretical Physics, Jagiellonian University, ul.~Łojasiewicza 11, 30-348 Kraków, Poland}

\begin{abstract}
Autoregressive Neural Networks (ANN) have been recently proposed as a mechanism to improve the efficiency of Monte Carlo algorithms for several spin systems. The idea relies on the fact that the total probability of a configuration can be factorized into conditional probabilities of each spin, which in turn can be approximated by a neural network. Once trained, the ANNs can be used to sample configurations from the approximated probability distribution and to explicitly evaluate this probability for a given configuration. It has also been observed that such conditional probabilities give access to information-theoretic observables such as mutual information or entanglement entropy.  In this paper, we describe the hierarchical autoregressive network (HAN) algorithm in three spatial dimensions and study its performance using the example of the Ising model. We compare HAN with three other autoregressive architectures and the classical Wolff cluster algorithm. 
Finally, we provide estimates of thermodynamic observables for the three-dimensional Ising model, such as entropy and free energy, in a range of temperatures across the phase transition.
\end{abstract}


\end{frontmatter}

\section{Introduction}
\label{sect_intr}

The recent idea \cite{PhysRevD.100.034515, 2020PhRvE.101b3304N} of employing artificial neural networks in Monte Carlo simulations has given birth to a new class of algorithms that promise to outperform traditional state-of-the-art algorithms, such as the cluster algorithm \cite{WOLFF1989379}. This objective remains a future goal because of the problematic scaling with the system size and difficulties in training the neural networks. Hence, the quest for more optimal neural network architectures, see for example \cite{Biazzo_2023,2024MLS&T...5b5074B, Singha:2025lsd}.  Besides the objective of improved efficiency, several works have observed that such new algorithms can provide access to observables that may be difficult to estimate in traditional Markov Chain Monte Carlo (MCMC). Examples of such observables include thermodynamic quantities such as the free energy, entropy  in lattice field theory \cite{2019PhRvL.122h0602W, Nicoli:2020njz} or information-theoretic observables such as mutual information \cite{Bialas:2023fjz} and quantum entanglement entropy \cite{Bialas:2024gha, Bialas:2025ldp}. So far, most applications have been performed for two-dimensional statistical systems or one-dimensional quantum systems, but higher-dimensional systems were also considered -- see {\it e.g.}~\cite{Abbott:2023thq}. 

The hierarchical algorithm proposed in Ref.~\cite{Bialas:2022qbs} uses a set of autoregressive neural networks \cite{2019PhRvL.122h0602W} that enable a recursive fixing of all spins on a square lattice. It was devised for two-dimensional spin systems and was benchmarked in the Ising model. The construction relies on the Hammersley-Clifford theorem \cite{Hammersley-Clifford, Clifford90markovrandom} and the Markov field property of the Ising model, i.e., as long as the interactions are nearest-neighbor only, for any closed contour, the probability of any spin inside the contour depends conditionally only on the values of spins from the boundary. However, the theorem underlying this idea is valid for systems in any number of dimensions; for example, in 3 dimensions (3D), a closed surface is needed. 

The goal of this work is to provide a generalization of the hierarchical algorithm to three dimensions and to benchmark its scaling with the system size and the training efficiency. We shall also compare it with architectures based on convolutional networks. Again, we demonstrate the algorithm using the example of the Ising model.

We consider $N=L^3$ spins that are located on a $L\times L \times L$ periodic lattice. The energy of a configuration of spins, $\mathbf{s} = \{ s^i \}_{i=1}^N$, is given by:
\begin{equation}
    E(\mathbf{s}) = - J \sum_{\langle i,j \rangle} s^i \, s^j \,,
\label{Ising_hamilt}
\end{equation}
where the sum runs over all neighboring pairs of lattice sites and $s^i=\pm 1$. We set the coupling constant $J=1$, and hence ferromagnetic interactions are considered. Then, each configuration $\mathbf{s}$ appears with the Boltzmann probability
\begin{align}
    p(\mathbf{s}) = \frac{1}{Z} \exp(-\beta E(\mathbf{s})) \,,
    \label{eq_boltz_distr}
\end{align}
where $\beta$ is an inverse temperature, $\beta=1/T$; $Z$ is the partition function, $Z=\sum_{\mathbf{s}} \exp(-\beta E(\mathbf{s}))$ and the sum is performed over all $2^N$ configurations.

\section{Neural samplers}

\subsection{Autoregressive networks}
The basic idea of Variational Autoregressive Networks (VAN)  \cite{2019PhRvL.122h0602W} is to represent the probability of a spin configuration (\ref{eq_boltz_distr}) as a product of $N-1$ conditional probabilities:
\begin{equation}\label{eq:conditional_probabilities}
    p(\mathbf{s}) = p(s^1)\prod_{i=2}^N p(s^i| s^1, \dots, s^{i-1}).
\end{equation}

The analytic form of conditional probabilities $p(s^i|\cdot)$ is not known; however, one can use neural networks to approximate them. We denote by $q_{\theta}(s^i|\cdot)$ the neural approximation, where the subscript $\theta$ indicates the weights of the neural networks. Then,
\begin{equation}\label{eq:conditional_probabilities}
    q_{\theta}(\mathbf{s}) =  q_{\theta}(s^1)\prod_{i=2}^N q_{\theta}(s^i| s^1, \dots, s^{i-1}),
\end{equation}
is the probability distribution of spin configurations provided by the network. There are two crucial properties of the algorithm. First, the network enables a direct sampling of the probability distribution $q_{\theta}$, which is not possible for the target distribution $p$. Second, network training enables $q_\theta$ to be a good approximation of $p$.\footnote{ Note that there are other neural samplers, which possess those important properties, {\it e.g.}, normalizing flows or diffusion models.}

In the VAN algorithm \cite{2019PhRvL.122h0602W}, the generation of spin configurations is performed using a single neural network that takes as input the values of the consecutive spins and gives at the output the probabilities of the spins to have value $+1$. The network is evaluated $N$ times and at $i$-th evaluation the (conditional) probability distribution of the $i$-th spin, $q_{\theta}(s^i| s^1, \dots, s^{i-1})$, is calculated based on the values of the $i-1$ spins fixed in the previous steps. Then, the $i$-th spin is fixed using its probability distribution. In principle, such an algorithm can sample any spin system in arbitrary dimensions with any type of interaction.

The architecture of the network needs to follow the functional dependencies of the conditional probabilities, namely $i$-th network's output must depend only on the first $i-1$ inputs (hence the name "autoregressive"). To realize this in practice, we rely either on dense layers (which is known as the Masked Autoencoder for Distribution Estimation (MADE) \cite{2015arXiv150203509G}) or on convolutional networks. In both cases, some of the weights are multiplied by zero to "remove" unwanted functional dependencies (see Section \ref{section_gated_CNNs} for more details).

The ability of the VAN algorithm to sample from the distribution $q_\theta$, where the probability of a spin configuration is explicitly calculated, allows for the so-called reverse training using the (reverse) Kullback–Leibler (KL) divergence,
\begin{equation}\label{eq-KL}
    D_{KL}(q_{\theta}|p) = \sum_{\mathbf{s}} \, q_{\theta}(\mathbf{s}) \log \frac{q_{\theta}(\mathbf{s})}{p(\mathbf{s})} = \avg{\log q_{\theta}(\mathbf{s})-\log
p(\mathbf{s})}_{q_{\theta}},
\end{equation}
 where $\avg{\ldots}_{q_{\theta}} := \sum_{\mathbf{s}} q_\theta(\mathbf{s}) (\ldots)$ is a statistical average over the distribution $q_\theta$.
The KL divergence measures the difference between two probability distributions $q$ and $p$: $D_{KL}(q | p) \geq 0$ and $D_{KL}(q | p) = 0 \Leftrightarrow q = p$ (note, however, that in general $D_{KL}(q | p) \neq D_{KL}(p |q  )$ so $D_{KL}$ is not a distance in the mathematical sense).  In a typical situation in Machine Learning, where data are easily available, one usually applies forward training, where $D_{KL}(p |q_\theta)$ is minimized. This is also possible in the context of statistical physics or lattice field theory; however, it requires configurations generated from $p$ using, for example, some MCMC algorithm. One estimates the average on the r.h.s. of \eqref{eq-KL} using a batch of $n$ samples drawn from distribution $q_\theta$:
\begin{equation}
    \avg{\ldots}_{q_{\theta}} \to \frac{1}{n} \sum_{i=1}^n (\ldots), \textrm{ where } \  \mathbf{s}_i \sim q_{\theta}.
\end{equation}
If $p$ is given by (\ref{eq_boltz_distr}), one defines:

\begin{equation}\label{eq:Fq-autoregressive}
    \begin{split}
    \beta \hat F_q =\frac{1}{n} \sum_{i=1}^n \left(\log q_{\theta}(\mathbf{s}_i)+\beta E(\mathbf{s}_i) \right),\quad \mathbf{s}_i \sim q_{\theta},
\end{split}
\end{equation}
which is minimized during the network training. Minimizing $\beta \hat F_q$ is equivalent to minimizing (\ref{eq-KL}), as 
$D_{KL}(q_{\theta}|p)=\beta F_q - \beta F$, where $F_q= \avg{\log q_{\theta}+\beta E}_{q_{\theta}}/\beta$  
and the additive constant is the free energy:
\begin{equation}
F= -\frac{1}{\beta} \log Z.
\label{F_def_eq}
\end{equation}

\subsection{Neural importance sampling (NIS)}

In practice, there is always some difference between $p$ and $q_\theta$ as the networks cannot be perfectly trained. This difference may introduce a bias when the mean values of the observables are calculated. In this manuscript, we shall use the so-called Neural importance sampling (NIS) \cite{2020PhRvE.101b3304N} method to get unbiased results, which is based on the reweighting procedure.\footnote{In the alternative approach, Neural Markov Chain Monte Carlo (NMCMC) \cite{PhysRevD.100.034515,2020PhRvE.101b3304N}, one uses the accept-reject step on the configurations generated by the network.}  For some observable $\mathcal{O}(\mathbf{s})$, the expectation value is given by:
\begin{equation}
    \avg{ \mathcal{O} } = \frac{1}{{\cal N}\hat{Z}} \sum_{i=1}^{\cal N}  \hat w(\mathbf{s}_i) \mathcal{O}(\mathbf{s}_i), \qquad \mathbf{s}_i \sim q_{\theta},
    \label{nis_av_def}
\end{equation}
where the so-called importance ratios are defined as
\begin{equation}
\hat w(\mathbf{s}_i) = \frac{e^{-\beta E(\mathbf{s}_i)}}{q_\theta(\mathbf{s}_{i})},
\end{equation}
and the estimate of the partition function is given by:
\begin{equation}
    \hat{Z} = \frac{1}{{\cal N}}\sum_{i=1}^{\cal N} \hat w(\mathbf{s}_i), \qquad \mathbf{s}_i \sim q_{\theta}.
    \label{Zhat def}
\end{equation}

It is easy to see that when $q_{\theta}=p$ one has $\hat w(\mathbf{s}_i)=Z$ for all $\mathbf{s}_i$ and (\ref{nis_av_def}) reduces to the standard average. When $q_{\theta}\ne p$, the importance weights depend on the configuration, and the reweighting (\ref{nis_av_def}) compensates for the deviation between these two distributions. 

Two notes are in order here. First, it is crucial for the NIS method that $q_{\theta}(\mathbf{s}_i)$ is known\footnote{Some generative models can generate configurations but are unable to provide their probability, for instance Generative Adversarial Networks (GANs).} and is nonzero for all configurations in support of $p$. Second, the definition (\ref{Zhat def}) provides an unbiased estimator of the partition function $Z$; hence, NIS gives access to any thermodynamic observable that may be obtained from $Z$ -- see the end of section \ref{phys_obs_sect}. This is the most significant advantage of NIS compared to standard Monte Carlo algorithms, where getting $Z$ is cumbersome.

In the context of NIS, a useful measure of the difference between $q_\theta$ and $p$ is the Effective Sample Size ($ESS$)\cite{Liu, Kong}:
\begin{equation}
    ESS= \frac{\avg{\hat{w}}_{q_\theta}^2}{\avg{\hat{w}^2}_{q_\theta}} \approx \frac{\left(\sum_{i=1}^{\cal N} \hat w(\mathbf{s}_i)\right)^2}{{\cal N} \sum_{i=1}^{\cal N} \hat w^2(\mathbf{s}_i)},
\label{ESS_definition}
\end{equation}
where $\approx$ denotes the approximation of a finite batch size of samples drawn from $q_\theta$. 
Values of $ESS$ are in the range $(0,1]$ and for $q_\theta=p$ one has $ESS=1$.\footnote{Note that $ESS=1$ does not mean that $q_\theta=p$ since the mode collapse may occur, see discussion at the end of section \ref{phys_obs_sect}.} The ESS is a measure of how effective the sampler is, namely the statistical uncertainty of the observable obtained using NIS is $\sigma_{stat}\propto 1/\sqrt{ {\cal N}\cdot ESS}$, where $\cal N$ is a number of generated samples.

\section{Architectures of networks}
\label{archit_sect}

\subsection{Hierarchical autoregressive networks}
\label{han_section}

\begin{figure}[h]
    \begin{center}
    \includegraphics[width=0.3\textwidth]{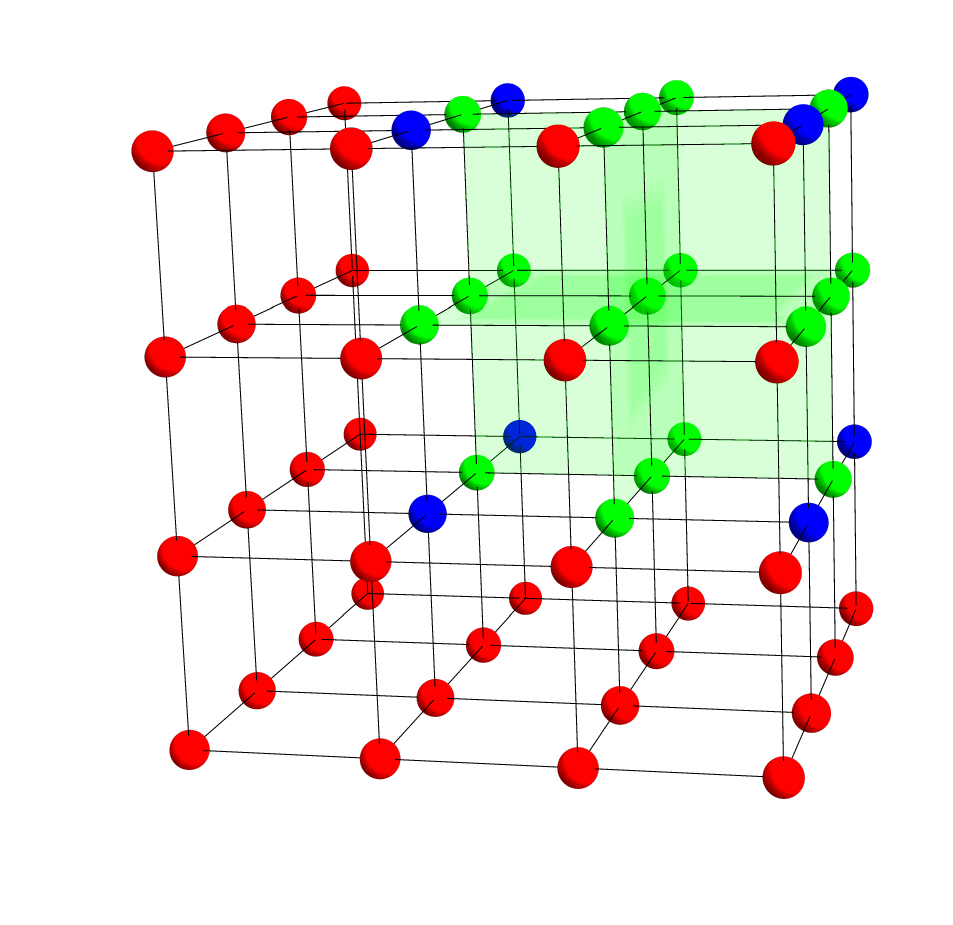}\hspace{3mm}%
    \includegraphics[width=0.15\textwidth]{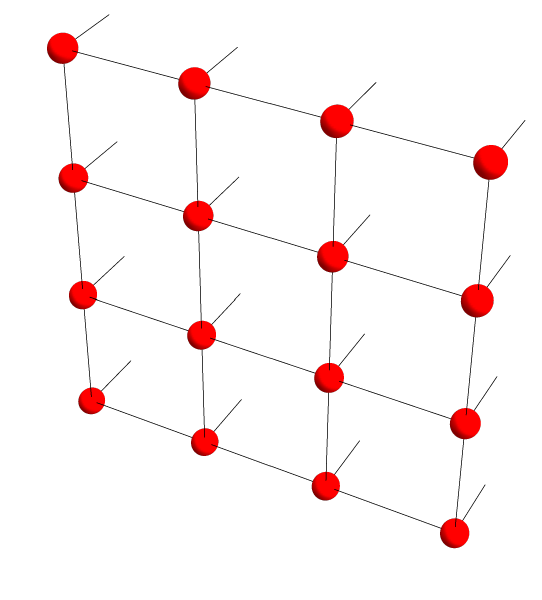}\hspace{3mm}%
    \includegraphics[width=0.15\textwidth]{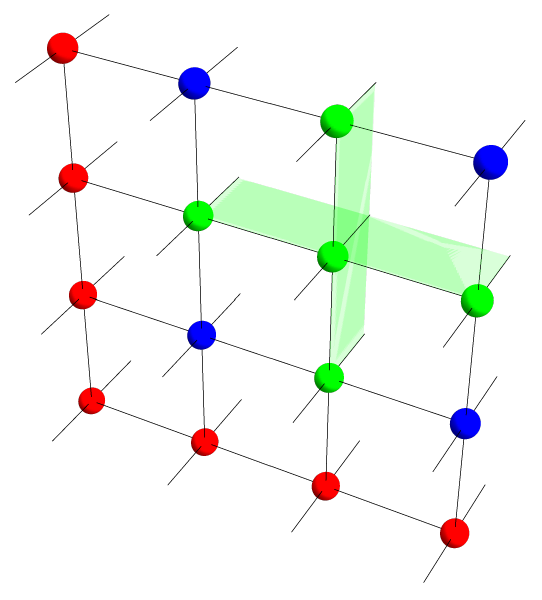}
    \includegraphics[width=0.15\textwidth]{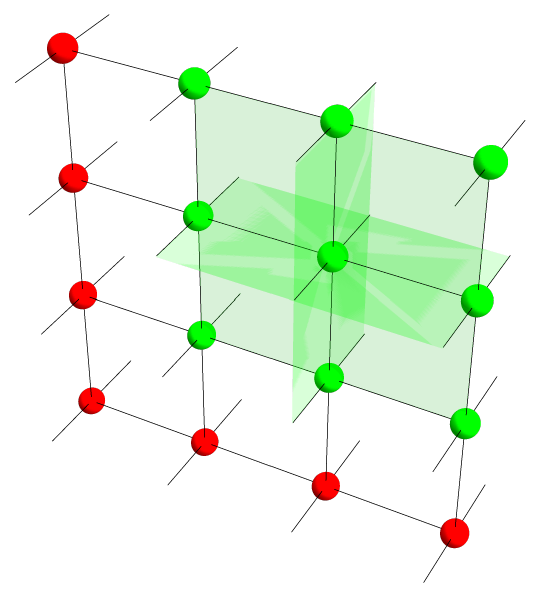}\hspace{3mm}%
    \includegraphics[width=0.15\textwidth]{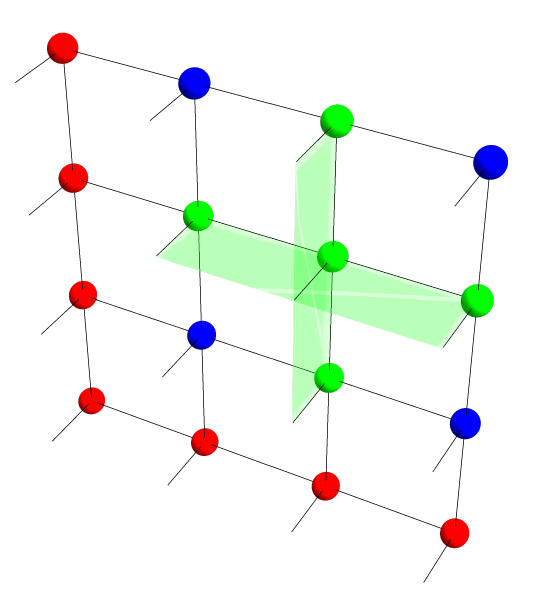}
    \end{center}
    \caption{Cube on the left part of the Figure: a hierarchical partitioning in 3D for $L=4$. The full system of $4\times4 \times4$ spins is divided into 3 subsets (denoted with the colors). Central and right part of the Figure: for better visibility of the 3D system, we also show four slices of the $4\times4 \times4$ cube. Spins generated with the networks at a given step of hierarchy are denoted with colors (see the text for details): red (first step and network), green (second step and network), blue (third step -- heat-bath algorithm).
    }
    \label{fig:sublattices4}
\end{figure}

In the VAN algorithm, it is necessary to evaluate the neural network $N$ times to fix all $N$ spins. Since all conditional probabilities (\ref{eq:conditional_probabilities}) are provided by a single network, this network must have $N$ input and $N$ output neurons. In the implementation of VAN based on dense neural networks (MADE), the number of weights of the network scales as $\sim N^2$. Assuming that the whole matrix multiplications are performed (which is usually true, due to the ease of implementation), the cost of generating one configuration scales as $\sim N^3$, which is $\sim L^9$ for 3D systems.  
 Note that this naive counting does not take into account the numerical cost of handling data, etc., which can significantly modify the scaling at small $L$.

The unfavorable scaling of the numerical cost of VAN, coming from the evaluation of a large network multiple times, is an inspiration for the development of an improved algorithm, called Hierarchical Autoregressive Networks (HAN). In HAN, one replaces a single neural network from the VAN approach with a set of smaller neural networks. In Fig.~\ref{fig:sublattices4} we show a schematic representation of the spin system of size $4\times 4 \times 4$ and demonstrate, using colors, its partition into recursive subsets (cube on the left). The right part of Fig.~\ref{fig:sublattices4} presents 4 slices of the 3D cube (presented for better visibility). Periodic boundary conditions are imposed, namely, spins in the outer wall interact with the spins in the opposite wall.

We start by explaining the HAN algorithm for the case $L=4$ (shown in Fig.~\ref{fig:sublattices4}) and later discuss larger systems. For $L=4$ HAN 3D consists of the following steps:
\begin{enumerate}
\item We start generation with the so-called boundary spins, denoted by red dots in Fig.~\ref{fig:sublattices4}. For this, we are using a standard autoregressive network - we call this the first step of the network's hierarchy. Due to periodic boundary conditions, these three walls provide the boundary that encloses the interior of the system (in Fig.~\ref{fig:sublattices4} interior is a $3\times 3 \times 3 $ cube of spins denoted with green and blue dots).

\item In the second stage of the hierarchy, the second network is used to fix the 3d cross-like structure, denoted by green dots in Fig.~\ref{fig:sublattices4}. The network takes as input not only the values of spins denoted as green dots but also the red ones. As the output, it provides conditional probabilities for the green spins. This dependence on the boundary spins was introduced by adding additional input to dense layers, keeping the autoregressive dependence of the generated spins, see Fig.~2 in Ref.~\cite{Bialas:2022qbs} for the schematic graph.

\item For $L=4$, the hierarchy ends with 8 blue spins, which are isolated (in the sense that all their neighbors' values are fixed). The isolated spin can be fixed using the heatbath algorithm, namely, one explicitly calculates the probability of the spin $s^i$ to be $+1$ from the expression:
\begin{equation}
    q(s^i= +1, n(s^i))=  \left[ 1+\exp \left(-\, 2\beta  \sum_{\ j \in n(s^i) } s^j \right) \right]^{-1}, 
    \label{heatbath_form}
\end{equation}
where the sum is performed over neighbors $n(s^i)$ of the spin $s^i$.
\end{enumerate}

For $L=8$ at the first stage of the hierarchy, one needs to fix spins on 3 walls, each of size $8 \times 8$, 169 spins in total, because some spins are shared between walls. In the second stage, the 3D cross-like structure has a size of 7 spins in every dimension and divides the spins into 8 cubes, each of size $3 \times 3 \times 3$. At the third step, we can use the same network architecture as that used in the $4 \times 4 \times 4$ system (green spins in Fig.~\ref{fig:sublattices4}).  Note that the conditional probabilities of those spins have the same functional dependencies on the surrounding spins - this is a consequence of the Hammersley-Clifford theorem (see Section \ref{sect_intr}). 
Therefore, the spins in all 8 cubes (of size $3 \times 3 \times 3$) can be fixed simultaneously using only one network.  In other words, at the third step of the hierarchy in $L=8$, we have one network that fixes the cross-like structures (green dots in Fig.~\ref{fig:sublattices4}) in eight cubes in parallel. The last stage of hierarchy is the isolated spins, which are fixed using the heat-bath step with probabilities (\ref{heatbath_form}).

The procedure for $L=16,32,64\ldots$ follows the same algorithm. Going from $k$ to $k+1$ (where $L=2^k$) requires: i) adding (on the second step) a new network and ii) increasing the size of the first network (first step), which fixes boundary spins. 
Snippets of the code that implement HAN 3D can be found in \ref{impl_details}.

Let us now calculate the total number of matrix multiplication operations that need to be performed to fix all the spins in the configuration using HAN. This would be asymptotically the leading computational cost of the algorithm.
We assume 2 layers at each level of hierarchy, as we used in our implementation. For simplicity, we double-count some spins: for example, at the first level (red spins in Figure \ref{fig:sublattices4}) we have $L^2+L(L-1)+(L-1)^2$ spins (3 walls), but we approximate this by $3 L^2$ - this will obviously not affect the scaling at large $L$. As usual, $L$ is a power of $2$, but is not smaller than 4.

The first network in the hierarchy has $3L^2$ inputs and $3L^2$ outputs. We run the network $3L^2$ times, as this is the number of spins at this level of the hierarchy. Each spin generation requires double (we assumed two layers) multiplication of a $3L^2\times 3L^2$ matrix by a vector. So in total, we have $2 \times 3L^2\times 3L^2$ multiplications per spin and $2 (3L^2)^3=54L^6$ operations in total.

At the second level of the hierarchy, one cross-like structure is fixed (green spins in Figure \ref{fig:sublattices4}), containing $3 L^2$ spins. At this level, the network has $3L^2+6L^2=9L^2$ inputs, where $6L^2$ is the number of spins surrounding the cross-like structure (in this case, it is the surface of the whole cube -- note the redundancy here: we have $3L^2$ independent spins, but we feed them "duplicated" to the network as $6L^2$ spins). The first layer has $9L^2$ inputs and $9L^2$ outputs, while the second layer has $9L^2$ inputs and $3L^2$ outputs. So, the number of operations at this level is $3L^2\times 9L^2 \times 9L^2 + 3L^2\times 9L^2 \times 3L^2 = 324 L^6$.

At the third level of the hierarchy, one fixes 8 cross-like structures, each of size $3(L/2)^2$ spins. Each of the cross-like structures has $6(L/2)^2$ spins surrounding it. For a single cross-like structure, we can use the result from the previous paragraph, replacing $L\to L/2$: so we have $324 (L/2)^6$ operations. For 8 cross-like structures, this would be $(324/8) L^6$ operations in total.

It is easy to see that the pattern will continue: going one step down the hierarchy, the number of operations gets a factor $1/8$ (the factor 8 due to more structures times the factor $1/64$ due to the smaller size of the cross-like structure). So, the number of multiplication operations from all the levels of the hierarchy for $L=2^k, k\ge 2$ will be:
$54L^6[1+6(1+ 8^{-1}+ 8^{-2} + 8^{-3} +\ldots + 8^{-(k-2)}) ] = 54L^6 \left[ 1+ \frac{48}{7}(1-8L^{-3}) \right]\approx (2970/7) L^6$, where in the approximation we skipped the sub-leading term since, from the beginning, we neglected them.\footnote{
The scaling with system size for the 3D hierarchical algorithm can also be determined using the master theorem for divide-and-conquer recurrences \cite{10.1145/1008861.1008865}. The problem has the following recurrence relation (for second and higher steps): $T(L) = 8T(L/2) + \Theta(L^{6})$, which 
after applying the master theorem gives the scaling of $\Theta(L^6)$. The first network is not a part of the recursion, but it also has a scaling $\Theta(L^6)$, so overall the scaling remains $\Theta(L^6)$.
}

For VAN, counting the number of multiplications is very simple: assuming two layers, we have two $ L^3 \times L^3$ matrices and the network needs to be evaluated $L^3$ times; so $2L^9$. Plugging in the value $L=8$ gives $2.7\times 10^8$ multiplications. This can be compared with the leading term for HAN, $(2970/7) 8^6\approx 1.1 \times 10^8$.

One should keep in mind that the scaling of $L^6$ in the number of multiplication operations does not automatically translate into the $L^6$ running time dependence, as other issues, e.g., parallelization on the graphics card, play an important role. Only in the limit of large $L$ we are expecting to see this behavior.

 The HAN algorithm can also be implemented for sizes $L$ that are not powers of $2$. There are several possible extensions of this kind. Probably, the simplest one is to replace the heat bath algorithm (at the last stage of the hierarchy) by yet another network that generates a cube of spins (e.g., of size $2\times 2 \times 2$). The same idea can be used to implement HAN for systems with cuboid shapes (non-cubic). We left those modifications for future work. 

In the simplest implementation we present here, the unwanted (due to the autoregressive property) connections in the dense network are masked (weights multiplied by 0), but still, the operation of multiplication is performed. One may think of a different implementation, where, for a given conditional probability, only the necessary weights get multiplied. This trick can reduce the scaling of VAN and HAN: from $\sim L^9$ to $\sim L^6$ (for VAN) and from $\sim L^6$ to $\sim L^4$ (for HAN). However, the gains are not immediately obvious, as they depend on the size of the problem and the capacity of the GPU. Such an implementation is in progress and will be presented elsewhere.

Another neural-network-based approach, the TwoBo \cite{2024MLS&T...5b5074B}, has scaling $\sim L^5$, which is lower than HAN (in the current implementation). The RiGCS approach presented in Ref.~\cite{Singha:2025lsd} was shown to scale better than HAN in 2D; however, its version in 3D is not available. The comparison of those algorithms with the ones presented here is an interesting future direction of research. We should note that in HAN, due to sampling the boundary spins first, it is relatively easy to implement different boundary conditions. This is an important property for calculating observables such as classical mutual information or entanglement entropy (see Ref.\cite{Bialas:2023fjz, Bialas:2024gha, Bialas:2025ldp}). To our knowledge, neither TwoBo nor RiGCS allows for such flexibility.

The results presented below were obtained with the same hyperparameters for the VAN and HAN networks: we always used dense networks with two layers: the first is followed by a PReLU activation function and the second by a sigmoid. The ADAM optimizer was chosen (see section \ref{sect_timings} for discussion of learning rate choice). Inverse temperature $\beta$ was increased  during training ($\beta$--annealing). We also used the $Z_2$ symmetry of the Ising model: $p(\mathbf s)=p(-\mathbf s)$: it allows for symmetrization of the probability distributions \cite{2019PhRvL.122h0602W}: $q_\theta(\mathbf s) \to [q_\theta(\mathbf s)+q_\theta(- \mathbf s)]/2 \equiv \bar q(\mathbf s)$. The batch size used to calculate the loss function and update the weights was 1000 configurations (the update of weights is called epoch). We stopped training when $ESS$ stabilized.

\subsection{Pixel CNN VAN and Gated PixelCNN VAN}
\label{section_gated_CNNs}
In this section, we discuss the convolutional neural networks (CNN) version of the VAN \cite{2019PhRvL.122h0602W}. It is based on the PixelCNN network used to generate images \cite{oord2016conditional}. The autoregressive property is ensured by using a masked kernel (a kernel with some elements multiplied by zero).

\begin{figure}[h]
    \centering
\includegraphics[width=0.7\textwidth]{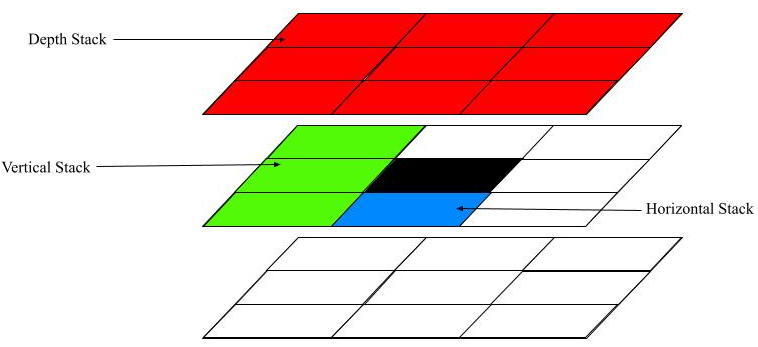}
\hspace{3mm}
    \caption{A 3D kernel of PixelCNN. The kernel is divided into stacks. Elements multiplied by 0 are denoted with white color.}
\label{cnn_kernel_fig}
\end{figure}
The original PixelCNN VAN \cite{2019PhRvL.122h0602W} was designed for 2D systems, so we upgraded the code to 3D. In Fig.~\ref{cnn_kernel_fig} we show a schematic view of the masked $3\times3\times3$ kernel. The red, green, and blue colors denote different parts of the kernel, called stacks. The white elements are multiplied by 0 (masked) so that the autoregressive property is preserved. The black element is either zero (which means a strictly autoregressive (s.a.) layer) or non-zero (weakly autoregressive layer).  In each autoregressive architecture, we use exactly one s.a. layer (the rest are weakly autoregressive) to ensure proper functional dependence of the conditional probabilities. In what follows, we shall call this architecture just {\em PixelCNN} since, except for autoregressive masking, it is based on standard 3D convolutional layers. 

\begin{figure}[h]
    \centering
\includegraphics[width=15cm,height=12cm]{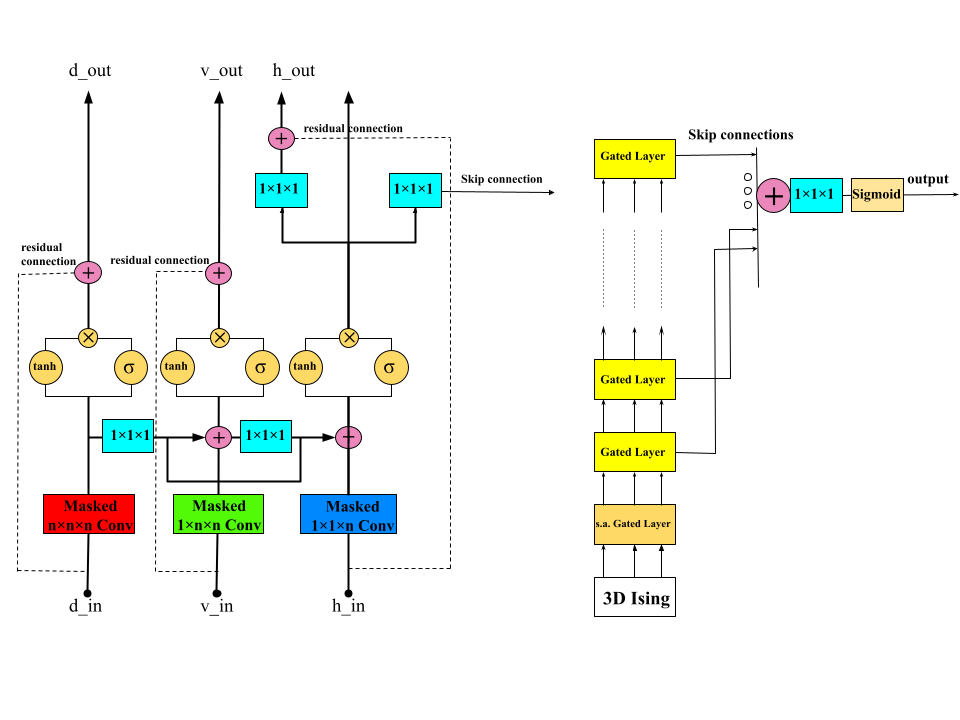}
    \caption{Left: schematic view of one gated PixellCNN layer. Right: architecture of gated PixelCNN network.  See text for the description.}
    \label{gated_cnn}
\end{figure}

In Reference \cite{oord2016conditional}, the Authors propose some improvements to the PixelCNN architecture. They note that the masking pattern leads to a blind spot in the receptive field even after applying many layers. To fix this, they propose to split the kernel into vertical and horizontal stacks. When adapting this architecture to 3D, we have added the third depth stack (see Fig.~\ref{cnn_kernel_fig}). The information coming from the values of spins in the depth stack (denoted as $\rm{d}_{\rm{in}}$) flows into both the vertical ($\rm{v}_{\rm{in}}$) and the horizontal stacks ($\rm{h}_{\rm{in}}$) and from the vertical to the horizontal stack using 1$\times$1$\times$1 convolutions (see Figure~\ref{gated_cnn} left).  To ensure that the model is autoregressive, the depth stack does not receive information from the vertical and horizontal stacks, and the vertical stack is blinded to information from the horizontal stack. The output of each stack is then fed into the Exponential Linear Unit (ELU) activation function. Finally, a residual connection is added to the output of each stack ($\rm{d}_{\rm{out}}$, $\rm{v}_{\rm{out}}$ and $\rm{h}_{\rm{out}}$) and a skip connection is drawn from the output of the horizontal stack. 
 In Ref.~\cite{oord2016conditional} the Authors propose also to use a gated architecture, which enhances the model performance. We shall follow this approach and call it {\em gated Pixel CNN}. 
The architecture of the gated 3D PixelCNN is shown in the right part of Fig.~\ref{gated_cnn}. The input is a 3D spin configuration, as for VAN/HAN architectures. The network consists of several gated layers, where the first one is strictly autoregressive (s.a.), and the rest are weakly autoregressive. The final output is the sum of all the skip connections coming from layers fed into the sigmoid function to get the conditional probabilities of the spins.

For convolutional networks, one expects $\sim L^6$ scaling with a system size: the number of convolutions needed to calculate a given conditional probability scales as $\sim L^3$; the number of spins scales as $\sim L^3$.

We used the same hyperparameters for PixelCNN and gated PixelCNN: networks consist of 6 layers, with 18 channels and kernel $5\times 5\times5$; the optimizer was ADAM and the learning rate was set to 0.003;  batch size was 1000. We stopped the training when $ESS$ stabilized.

\section{Numerical results}
\label{sec: results}

\subsection{Timings and ESS for autoregressive architectures}
\label{sect_timings}

 \begin{figure}[ht!]
     \centering
     \includegraphics[width=0.68\textwidth]{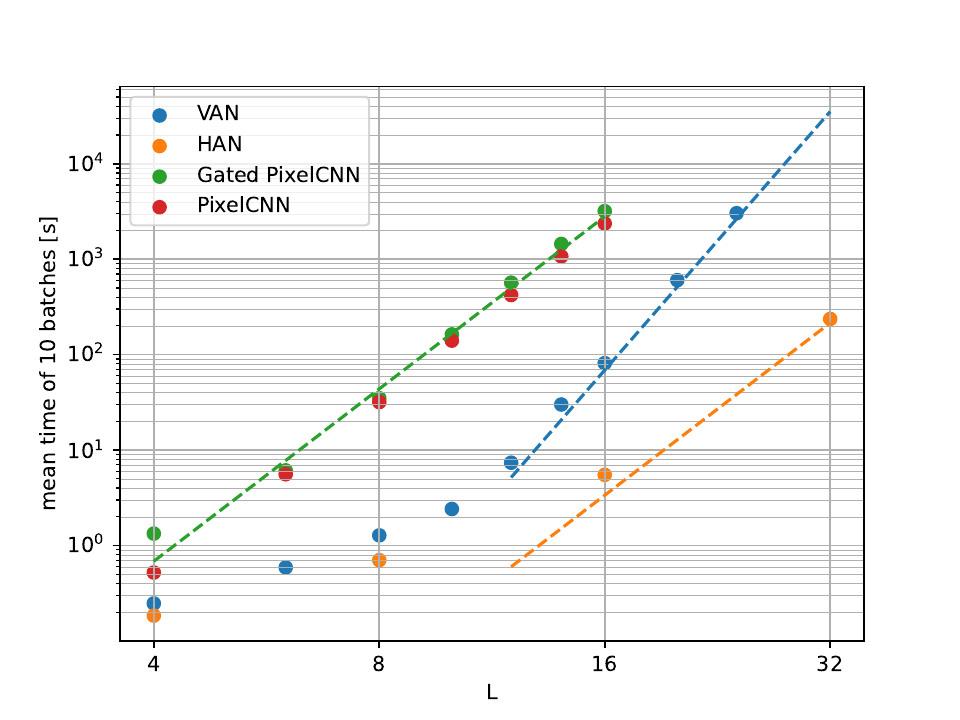}
     \caption{Mean time (in seconds) of 10 
     batches generation using the VAN (blue circles), HAN (orange circles), PixelCNN (red circles) and Gated PixelCNN (green circles) algorithms in dependence on system linear size $L$. Asymptotic approximations $\sim L^9$ (blue dashed line), $\sim L^6$ (orange dashed line) and $\sim L^6$ (green dashed line), for VAN, HAN and Gated PixelCNN accordingly, were superposed on plots. All the measurements were performed on Nvidia GeForce 4090 GPU.}
     \label{fig:speed}
 \end{figure}

We shall start by comparing different sampling algorithms when it comes to the configuration generation time \footnote{Since in autoregressive networks, numerical cost of training is dominated by the generation of the configurations, the conclusions of this analysis apply to training timings as well.}. We compare the generation time of 10 batches of $1024$ configurations using the architectures described in Section \ref{archit_sect}. In Fig.~\ref{fig:speed} we show the dependence on the linear system size $L$ on a log-log scale. 
First, we note that CNN-based architectures are significantly slower (an order of magnitude in the presented range of system sizes) than the VAN architecture based on dense networks. The HAN architecture is faster than VAN, as expected, and they have different scaling with system size, as we discussed in section \ref{han_section}. To better visualize the scaling of the architectures, we have added the lines in Fig. \ref{fig:speed}, which represent scaling $\sim L^9$ (blue line) and $\sim L^6$ (yellow and green line). For small system sizes, the scaling is modified by other effects ({\it e.g.} saving data or parallelization), but for larger $L$ we observe the expected scaling for all algorithms except HAN. Also, the timing scaling of CNN architectures follows the expected $\sim L^6$ behavior.

\begin{figure}
    \centering
    \includegraphics[width=0.49\linewidth]{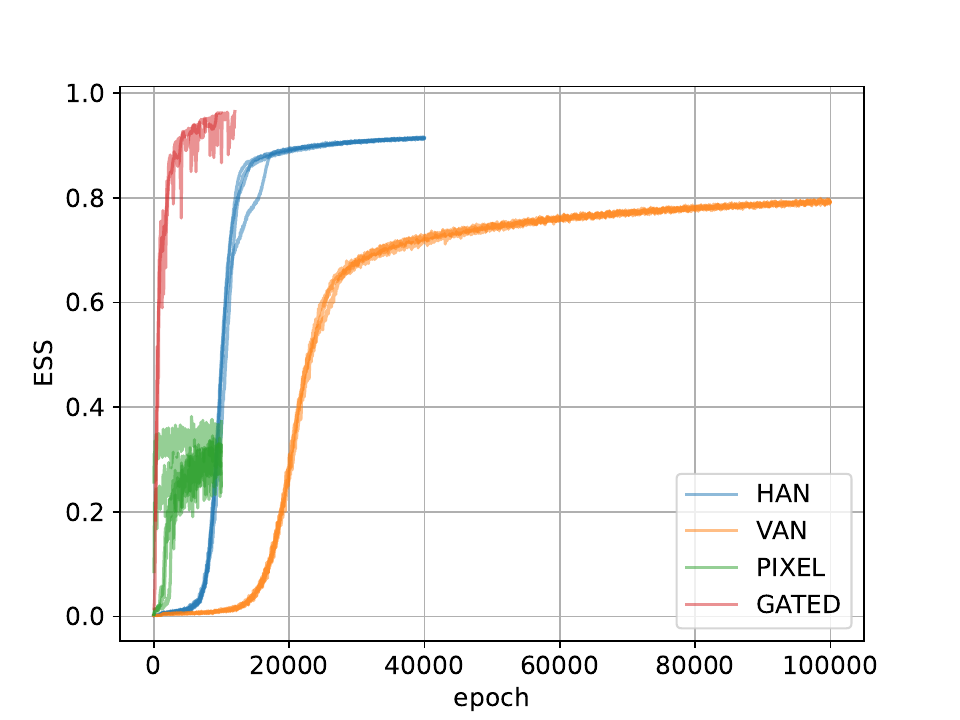}
    \centering
    \includegraphics[width=0.49\linewidth]{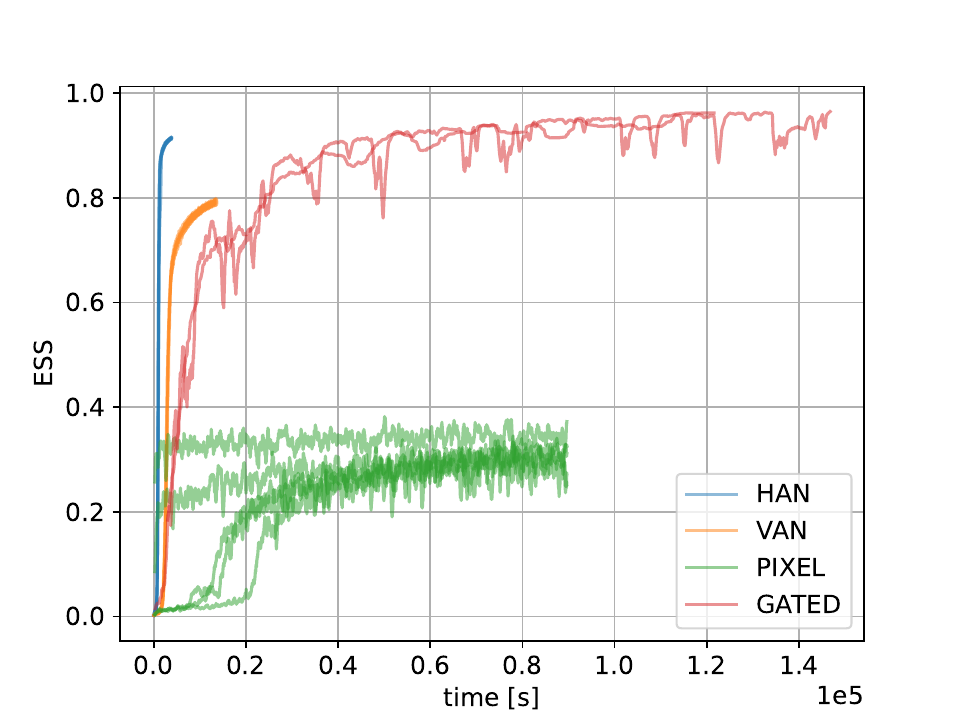}
    \caption{ Moving average of $ESS$ for different models as a function of epoch (left panel) and time (right panel) for $\beta = 0.22$, $L=8$. Curves with the same colors show different runs. }
    \label{fig:ess_models}
\end{figure}

To compare the training efficiency between different methods, we present in Figure \ref{fig:ess_models} how $ESS$ changes during their training for $L=8$ and $\beta=0.22$ (which is close to the critical temperature for the 3D Ising model -- see next subsection). We perform measurements of the $ESS$ at each batch of configurations that were used to calculate the loss function. Since the batch is small ($\sim1000$ samples), the statistical fluctuations of $ESS$ can be large; therefore, we perform a simple moving average over 50 epochs to get smoother curves. We ran the algorithms several times to check if the results are stable (group of curves with the same colors). For the purpose of this comparison, we used one learning rate (lr) for VAN and HAN, lr=0.0003. Note that usually, to get the best performance, one tunes the lr for each architecture separately. However, then the comparison of how $ESS$ changes with epochs is difficult to interpret. In the Pixel CNN and the gated Pixel CNN networks, we used lr=0.003, which was checked to be optimal for them - using the same lr as for dense networks would significantly slow down the training. On the other hand, for dense architectures, lr=0.003 is too large and the final $ESS$ is smaller than for lr=0.0003. The total number of epochs for each architecture was chosen such that the $ESS$ reaches its maximum. 

In the left panel of Figure \ref{fig:ess_models}  we show the $ESS$ as a function of the number of epochs. We see that CNN architectures (pixel CNN and gated pixel CNN) need fewer epochs to be trained than dense architectures (VAN and HAN). This is mostly due to the fact that lr for those architectures can be 10 times larger. One can also see that HAN needs $\sim 2$ times fewer epochs than VAN and reaches a better final $ESS$.
The situation is quite different if we plot the $ESS$ as a function of the time of training (right panel of Figure \ref{fig:ess_models} ). Since the generation of configurations in the gated Pixel CNN is much slower than in HAN, the HAN algorithm reaches its maximal $ESS$ in a much shorter time. We also notice that for Pixel CNN, the curve of training depends very much on the run; nevertheless, the final $ESS$ is always much lower than the values obtained with other models.

\begin{table}
    \centering
    \begin{tabular}{ccccc}
        System size & VAN & HAN & PixelCNN &  Gated PixelCNN\\
        4 & 95s & 26s & 84s & 38s \\
        6 & 794s & - & not reached & 991s \\
        8 & 4689s & 1078s  & not reached & 18668s\\
    \end{tabular}
    \caption{Time of training needed for moving average ESS to reach $0.7$ for $\beta = 0.22$, see the text for more details. The values were shown for one run, they can vary up to 10\% for different runs.}
    \label{tab:times_training}
\end{table}

We shall now demonstrate how the training time grows with the system size. For this purpose, we measure the time of a given architecture's training to the value of $ESS=0.7$.\footnote{The statistical error of the observables calculated using NIS is proportional to $1/\sqrt{ESS}$, therefore, the value of 0.7 leads to $20\%$ larger error than for a perfect sampler ($q_\theta=p$) -- for this comparison we assume it is a sufficient level to say that the network is well-trained.} In Table \ref{tab:times_training} we
show the timings obtained for $\beta=0.22$ and for three system sizes: $L=4,6,8$. Since the moving average of ESS has some statistical fluctuation, the value presented is the mean between the first time the threshold value 0.7  was reached by the moving average and the last time when the $ESS$ was below that value. The learning rates were the same for all system sizes and were chosen as described in the paragraph above (lr=0.0003 for VAN/HAN and lr=0.003 for PixelCNN/Gated Pixel CNN). 

 First, we observe that for all architectures the time of training grows much faster than the time of generation of batch presented in Figure \ref{fig:speed} (note that in the range $L \in [4,8]$ VAN and HAN do not follow their asymptotic scalings yet). This is just due to the fact that when increasing the size of the system, the networks need more epochs to be trained. As was shown for normalizing flow architectures in 2D, the numerical cost of training may scale exponentially with system size \cite{DelDebbio:2021qwf}. 
 Here, we cannot confirm such behavior as larger system sizes in 3D are currently out of reach due to small ESS values obtained. 
 
 For VAN, we observe a factor of $\sim50$ increase in training time by increasing the linear size of the system by a factor of 2; for HAN, this factor is $\sim40$. The fastest growth of training time is observed for Gated Pixel CNN, where the timing grows by a factor of $\sim 500$ between $L=4$ and $L=8$. Finally, the standard Pixel CNN architecture did not reach ESS=0.7 when applied to $L\ge 6$.

To check the maximal $ESS$ the networks can reach at $\beta=0.22$, we tune the lr for each architecture separately. For VAN and HAN, the lr was changed during the training from 0.005 to 0.0001 manually (we decreased lr when no progress in $ESS$ was observed). Finally, the VAN reaches $ESS$ of 0.8 and HAN of 0.95.
(measured at large statistics). For Gated PixelCNN, $ESS$ is similar to that for HAN, reaching around 0.95. For the PixelCNN, this value is smaller, around 0.4.

Using the same statistics, the difference in statistical uncertainties between different methods is not large (for example, the error for PixelCNN is $\sqrt{0.95/0.4}\approx 1.5$ larger than for HAN). In fact, the main difference between the methods is in the generation time of a given statistic (presented in Fig.~\ref{fig:speed}). In what follows, we focus on the HAN algorithm as it reaches very good $ESS$ and is faster than other methods. However, we confirmed that all the methods give the same results for physical observables, within the uncertainties.

\subsection{Physical observables}
\label{phys_obs_sect}

We shall now cross-check the results obtained using HAN for some physical observables using MCMC. For this purpose, we use an in-house implementation of the Wolff cluster algorithm \cite{WOLFF1989379}.  It is one of the state-of-the-art algorithms used in simulations of the Ising model due to a significantly reduced (compared to local-update Monte Carlo like the Metropolis algorithm) problem of critical slowing down close to the phase transition. At each step, one builds a cluster of spins (where the probability of a spin to be attached to the cluster depends on $\beta$) and then the spins in the cluster are flipped. 
The HAN was trained to $ESS=0.95$ and then $20\times 10^6$ configurations were generated for each inverse temperature $\beta$. For the Markov Chain generated using the Wolff algorithm,  $20 \times 10^6$ configurations were obtained to calculate observables.

We start with calculating the mean absolute magnetization per spin, defined as:
\begin{equation}
   \frac{\avg{ \left| M \right|}}{N}= \avg{\frac{1}{N} \left| \sum_{i=1}^N s^i\right|}_{p}.
\end{equation}
In the limit of an infinite volume ($L\to \infty$), this quantity is an order parameter of the system with $\avg{ \left| M \right|}/N=0$ for $\beta \to 0$ (disordered phase, high temperature), $\avg{ \left| M \right|}/N=1$ for $\beta \to \infty$ (ordered phase, low temperature) and the phase transition occurs at $\beta_c \approx 0.2215$.

\begin{figure}[ht!]
     \centering
     \includegraphics[width=0.49\textwidth]{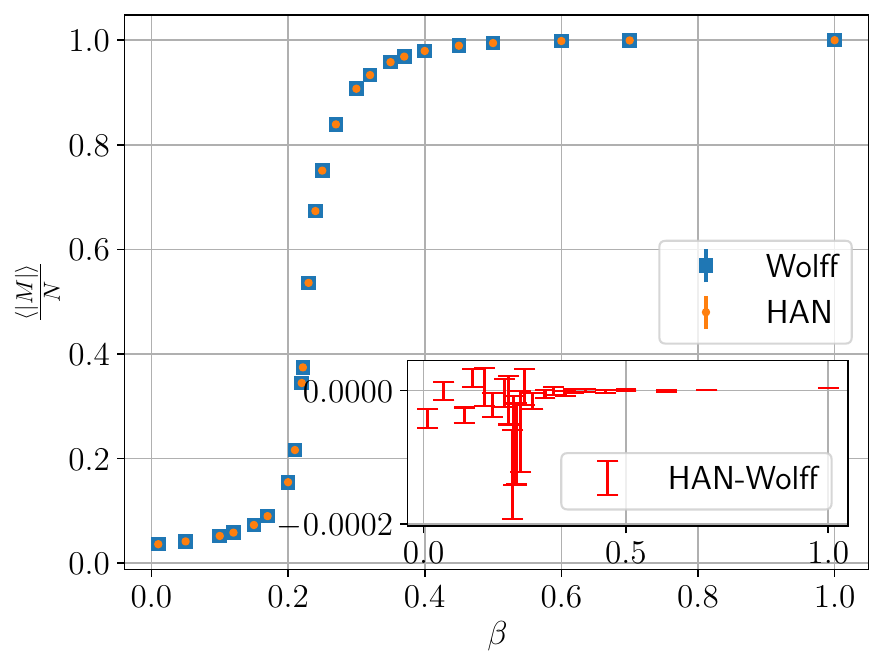}
     \includegraphics[width=0.49\textwidth]{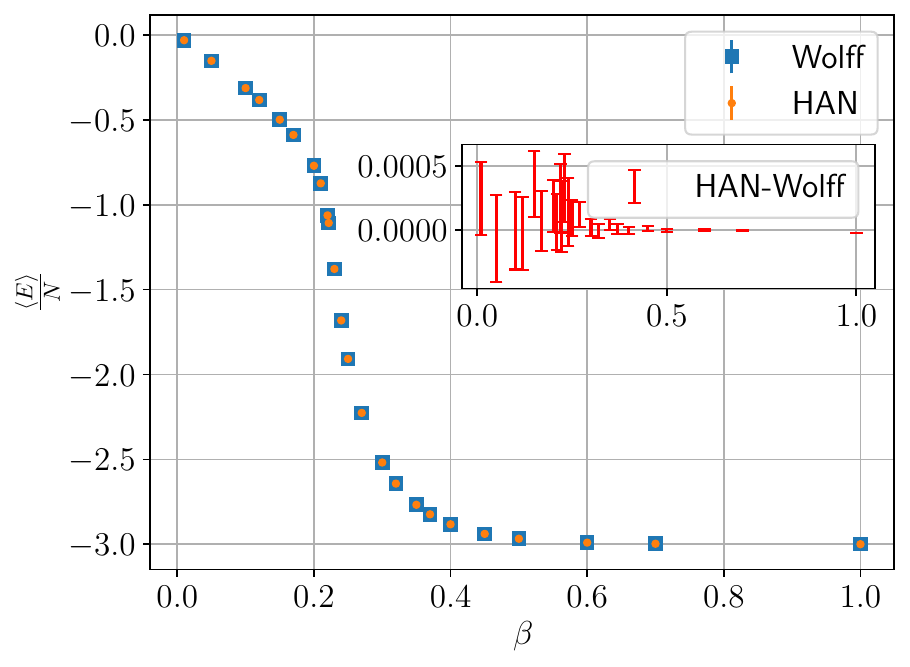}
     \caption{Comparison of the HAN (with NIS) with the Wolff algorithm for absolute magnetization per spin (left panel) and energy per spin (right panel). Insets show the difference between the two results; its uncertainty was obtained by adding the Wolff and HAN errors in quadrature.}
     \label{fig:mag_han}
 \end{figure}

In Fig.~\ref{fig:mag_han} (left), we show $\avg{ \left| M \right|}/N$ as a function of $\beta$ calculated for $L=8$ using two methods: the HAN (with NIS) and the Wolff algorithm. One can see very good agreement with the errors below 1\textperthousand, where the errors were obtained using the bootstrap method -- see the inset which shows the difference of the two results. In the right panel of Figure \ref{fig:mag_han}, we show the mean energy per spin $\avg{E}/N$, where a similar agreement can be observed. 

\begin{figure}[ht!]
     \centering
     \includegraphics[width=0.49\textwidth]{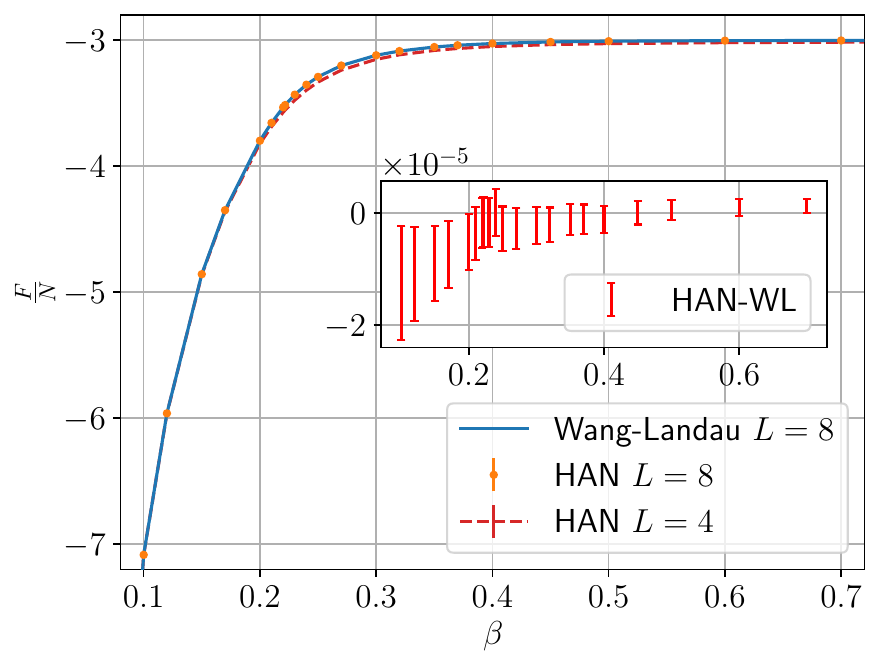}
     \includegraphics[width=0.49\textwidth]{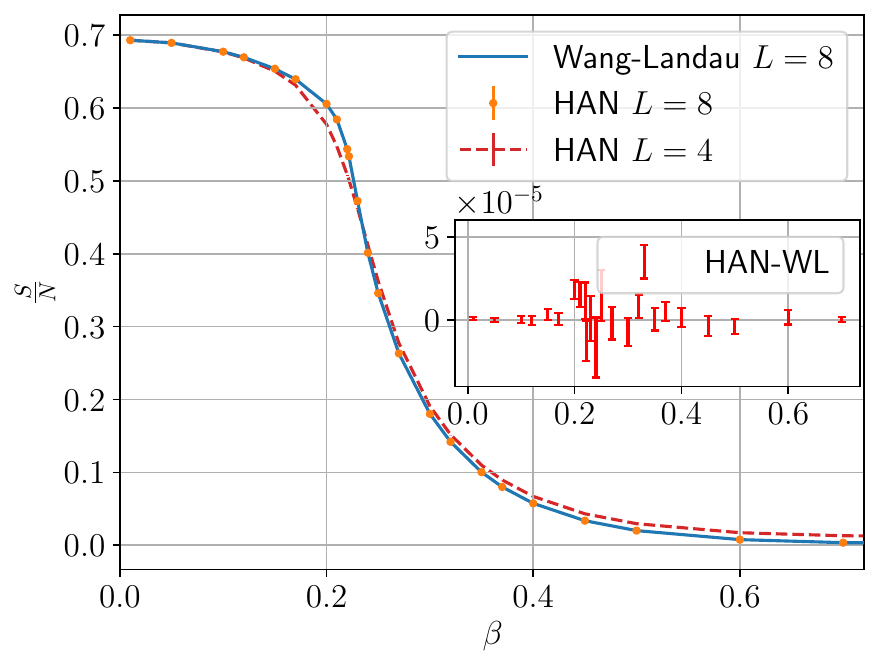}
     \caption{Free energy per spin (left) and entropy per spin (right) as a function of inverse temperature $\beta$. Results obtained using HAN (with NIS) were shown for two system sizes: $L=4$ (red dashed curve) and $L=8$ (orange points). For $L=8$, results from the Wang-Landau algorithm \cite{PhysRevLett.86.2050} were presented (blue solid curve). Insets show the difference between the two results; its uncertainty was obtained by adding the W-L and HAN errors in quadrature. 
     }
     \label{fig:f_s_han}
 \end{figure}

  We now move to the thermodynamical observables. In Figure \ref{fig:f_s_han}, we show results for the free energy per spin $F/N$ (\ref{F_def_eq}) and entropy per spin:
\begin{equation}
    \frac{S}{N}= \frac{\beta}{N} \left( \avg{E} -  F \right),
\end{equation}
where both quantities were calculated using the NIS method. We show results for $L=4$ (red dashed curve) and $L=8$ (orange points), demonstrating a weak dependence on the size of the system. To cross-check our results, we implemented the Wang-Landau algorithm (W-L)  \cite{PhysRevLett.86.2050}, which is a random walk algorithm capable of estimating the density of states. In Figure \ref{fig:f_s_han}, we show the results of W-L for L=8 with a solid blue line. The agreement between HAN and W-L is very good (see insets).

The results for W-L were obtained by averaging 1000 independent runs. In each run, the parameter of W-L $\ln f$ was halved when the flatness of the histogram reached 95\%.  The runs were started with $\ln f=1$ and stopped when $\ln f < 10^{-9}$. We used a dual-socket AMD EPYC 7F52 CPU (64 hardware threads in total) and the data presented in Figure \ref{fig:f_s_han} were calculated within $\sim 10^{4}$s.

 The HAN $L=8$ results from Figure \ref{fig:f_s_han} were obtained after $\sim 1000$s of training and $\sim 100$s of sample generation at each $\beta$ value; a GeForce RTX 4090 GPU was used. At $\beta=0.22$, the W-L reaches a similar statistical error (as HAN) for $S/N$ in $\sim 3000s$. However, to get similar errors for $F/N$, one would need to run the W-L for $\sim 10^{6}$s because the uncertainty of HAN for $F/N$ is much lower than for $S/N$. Although HAN is significantly faster, one should note that this is a very rough comparison; e.g., due to different hardware used to run both algorithms and a possible choice of a more optimal schedule for W-L. Also, in W-L, once the density of states is obtained, calculating the results for all temperatures is instantaneous, whereas HAN requires training at each temperature.

As stated above, NIS requires that the supports of $\qt$ and $p$ coincide. If they differ, we are talking about {\em mode collapse}, which is most often caused by broken symmetries (see \cite{nicoli2023modecolapse} for the discussion). Mode collapse can occur in reverse training, since the network generates its own training data (spin configurations). When a mode is missing in $q_\theta$, the $ESS$ can still be close to 1 because the configurations where $q_\theta$ seriously underestimates $p$ will never appear in the sampling.

We did not observe this mode-seeking behavior for the networks we trained; this is due to the $Z_2$ symmetry used in the training (see Section \ref{han_section}). 
In the absence of mode collapse, the NIS estimator of $Z$ given by equation \eqref{Zhat def} is unbiased. However, it has a non-zero variance ($ESS\neq 1$) due to imperfect training of the network. Because of this variance, estimators of other quantities that are functions of $Z$, like the free energy $F$ or entropy $S$, will usually be biased - see the Supplementary Materials of Ref.~\cite{PhysRevLett.126.032001} or the Appendix of Ref.~\cite{Bialas:2023fjz}. However, as was shown in Ref.~\cite{PhysRevLett.126.032001}, the bias decreases with statistics as $1/\cal N$; hence, for large statistics (and good $ESS$) the bias is negligible compared to the statistical error ($\propto 1/\sqrt{{\cal N}}$). To conclude, the systematic errors of NIS in the simulations we performed are insignificant.

\subsection{HAN versus Wolff algorithm: timings}
It is instructive to compare the speed of our HAN 3D algorithm with that of the Wolff cluster algorithm. Here, we are not aiming for a precise comparison, since we used the in-house code (with some optimizations) for the Wolff algorithm.  We shall make this comparison just to give the Reader a rough idea of HAN timings. The exact comparison of algorithms is difficult, also due to their very different nature. For example, the Wolff algorithm, in principle, is sequential; however, one can still run multiple Markov Chains simultaneously on the GPU. Both the Wolff algorithm and HAN were run on the same device: a high-performance Nvidia GeForce 4090 GPU.

The training of our HAN model at $\beta=0.264$ and $L=8$ took $\sim 430$s and reached $ESS=0.968$. With this model, we generated $~5\times 10^8$ configurations and, using NIS, calculated the relative error of the absolute magnetization equal to $2.6\times 10^{-6}$. The generation of configurations took $\sim 3400$s. To reach the same precision using the Wolff algorithm, one needs $~1.2\times 10^9$ configurations, where the autocorrelation time was measured to be 2.6. It takes $\sim 47$ s to generate them using 32768 parallel chains. 

In summary, the same precision for the observable was obtained with the cluster algorithm $\sim 70$ times faster than with our HAN 3D model (training and generation). It is worth noting that running the Wolff algorithm on a single CPU core is a few hundred times slower than using a GPU in our case.
We could use this "embarrassingly parallel" implementation because of a very small (compared to the number of collected samples) thermalization time, \textit{i.e.} time needed to reach statistical equilibrium. Were this time longer, it would add an overhead that could substantially reduce the advantage of using parallelization on a slower machine. Also, the size of the systems that can be simulated in this way is constrained by the memory of the GPU.

\subsection{Efficiency of training in 2D and 3D}

Here we compare how the HAN algorithm works in 2D and 3D systems. For this purpose, we consider a 2D system $8\times 8$ at $\beta=0.38$ and a 3D system $4\times 4\times4$ at $\beta=0.22$. The temperatures were chosen such that the magnetization is around 0.5 in both cases. The number of spins is the same in both systems, so the difference in training comes from the different setup of networks \footnote{HAN 2D in $8\times 8$ system uses 3 networks and HAN 3D in $4\times 4\times4$ system uses 2 networks.} and different correlations between spins in 2D and 3D. 

\begin{figure}[ht!]
     \centering
     \includegraphics[width=0.65\textwidth]{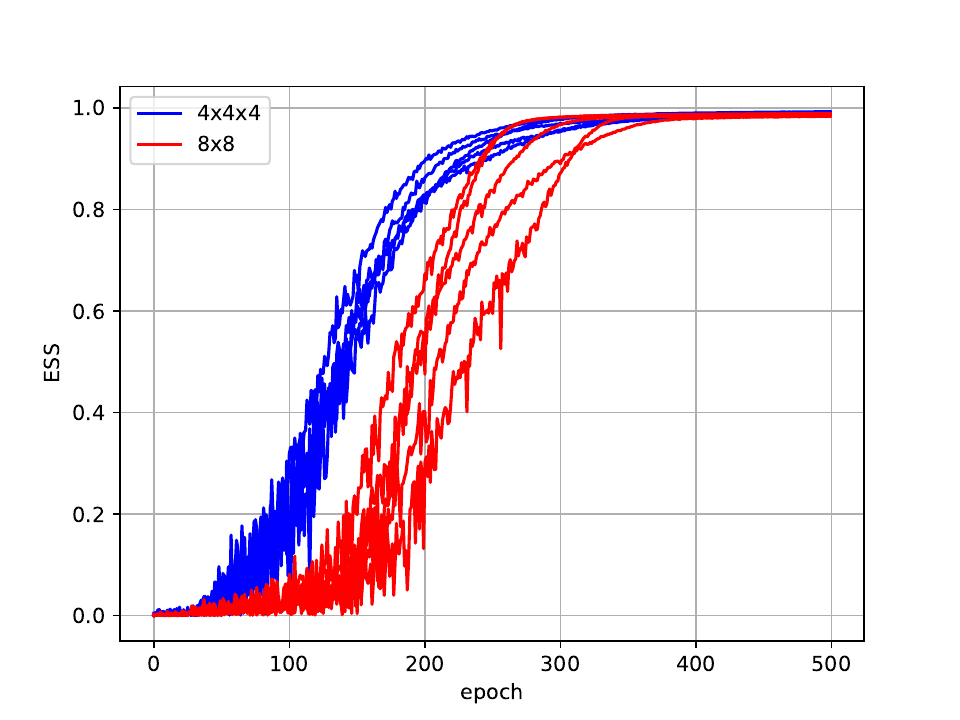}
     \caption{Comparison of the ESS growth for 4x4x4 (blue) and 8x8 (red) systems in a corresponding critical temperature (0.22 for the 3D and 0.38 for the 2D system) with the learning rate set to 0.005 and the batch size of 8196}
     \label{fig:esshan}
 \end{figure}

In Figure \ref{fig:esshan}, we plot the ESS as a function of training epochs for those two versions of the HAN algorithm. We used the same hyperparameters for dense networks (2 layers) and the same learning rate of 0.005. For each system, we run the training 5 times (curves of the same color). One can observe that ESS grows faster for 3D HAN than in 2D. 
The timings are the following: 500 epochs of training the 3D HAN last around 8.4 seconds, and around 10 seconds for the 2D HAN.

\section{Summary}

In this paper, we have presented the hierarchical autoregressive network (HAN) algorithm suited for 3-dimensional spin systems and compared it with three other architectures: dense variational autoregressive networks (VAN), autoregressive convolutional networks (PixelCNN), and a gated version of autoregressive convolutional networks (gated PixelCNN). The latter was never used in the context of spin generation. Although the 3D HAN algorithm uses the same idea as its 2D version, the code required significant changes concerning data manipulation when generating the spin configuration.  

We have performed several tests concerning the efficiency of 3D algorithms using the classical Ising model: i) scaling of generation time with the system size, ii) comparison with the Wolff algorithm (evaluation of physical observables as well as the timings). We have found that HAN 3D is the fastest of all the neural architectures tested and allows good training quality (measured $ESS \approx 0.95$) for the system $8\times 8 \times 8$. A rough comparison shows that for such a system, the Wolff cluster algorithm is much faster (factor $\sim 70$) on the same device.
Using HAN (or other architectures considered in this work), it is straightforward to obtain thermodynamic observables such as free energy or entropy. We compared our results with the Wang-Landau algorithm, finding good agreement.

In our opinion, the fact that autoregressive networks can be used to calculate observables often hardly accessible using standard MCMC algorithms, like mutual information or entanglement entropy, should be the main reason for their development. This manuscript is a step towards developing methods to calculate such observables in 3D systems. Those quantities have been challenging to obtain using other methods, such as tensor networks or Monte Carlo, based on the Jarzynski equality \cite{Bulgarelli:2023ofi}. The main obstacle to using NIS in 3D (and 4D) is the scalability of the neural samplers with system size. Although HAN is a good first step in this direction, much improvement is needed. Such an improvement may come from new architectures developed for general-purpose Machine Learning.

\section*{Acknowledgments}
We gratefully acknowledge Polish high-performance computing infrastructure PLGrid (HPC Center: ACK Cyfronet AGH) for providing computer facilities and support within computational grant no. PLG/2023/016395.
T.S. and D.Z. kindly acknowledge the support of the Polish National Science Center (NCN) Grant No.\,2021/43/D/ST2/03375.  P.K. acknowledges the support of the Polish National Science Center (NCN) grant No. 2022/46/E/ST2/00346. The study was funded by "Research support module" as part of the "Excellence Initiative -- Research University" program at the Jagiellonian University in Kraków. This research was partially funded by the Priority Research Area Digiworld under the program Excellence Initiative – Research University at the Jagiellonian University in Kraków.

\appendix

\section{Implementation details}
\label{impl_details}

In the Listing \ref{lst:draw_sample}, we show the snippet of the main function used for sampling the spin configurations in HAN 3D -- \verb|draw_sample|. With some small modifications ({\it e.g.}  dimension of tensors), it can be used for sampling spins in any number of dimensions. The function follows the points 1-3.~from section \ref{han_section}.

The \verb|draw_sample| function calls several functions that are implemented explicitly for three dimensions. For example, \verb|find_borders| (Listing \ref{lst:borders}) takes as input a cube of spins and returns the values of spins on its walls (but excluding the edges). The function \verb|make_cross| (Listing \ref{lst:cross}) takes as input the spin values and forms from them the 3D cross-like structure (see green spins in Figure \ref{fig:sublattices4}) - the output of the function is a cube with zeros everywhere except for the cross-like structure where $\pm1$ values are substituted. The function \verb|divide_into_cubes| (Listing \ref{lst:divide}) takes as an input a cube and returns a list of eight cubes with edges twice shorter than the original ones. There are several other functions used in the algorithm that need to be implemented for each dimension separately, unless some general versions of these functions are developed.

\begin{lstlisting}[float, caption={The main function used to sample configurations with HAN 3D. It creates a batch of 3D configurations of $L^3$ size each. Apart from the configurations, it returns the log of the probability, but only of the spins generated using the heat bath. This, together with a list of inputs to all the networks, enables us to compute the probability of the whole configuration. The heat bath generation on the last level of the hierarchy is done by the \texttt{heatbath} function. The hierarchical generation is handled by the \texttt{han} function described later. This function splits its input into small ($3^3$) cubes that have to be reassembled later by successive applications of the \texttt{add\_into\_cube} function.}, label=lst:draw_sample]
def draw_sample(
        net_b, int_nets,
        beta: float, L: int, batch_size: int,
        device: torch.device, float_dtype=torch.float32):
        
    # Sampling the three faces of the original lattice
    sample_b = net_b.sample(batch_size)
    list_args_for_nets: list[list[torch.Tensor]] = [sample_b]
    sample: torch.Tensor = create_cube(sample_b, L, L, L, device) 
    # sample is now a (L+1)^3 cube containing the faces of the original lattice
    # taken from sample_b, with opposite faces being duplicates of each other.

    sample, list_args_for_nets = han(sample, 
        int_nets, list_args_for_nets, device)
    
    # heat bath
    border = find_borders(sample)
    corona = torch.sum(border, dim=(2, 3))
    spins, log_prob_heatbath = heatbath(corona, beta, float_dtype)
    sample[:, 0, 1, 1, 1] = spins[:, 0]

    # recreation of original cubic sized sample 
    # from its subdivision into small cubes. 
    log_prob_heatbath = log_prob_heatbath.view(-1, batch_size).sum(0)
    while sample.shape[0] > batch_size:
        sample = add_into_cube(
        torch.chunk(sample, 8, dim=0))
        
    # delete the last faces, as they are duplicates of the first.
    sample = sample[:, :, :-1, :-1, :-1]
    
    return sample, list_args_for_nets, log_prob_heatbath
\end{lstlisting}

\begin{lstlisting}[float,caption={The \texttt{han} function that performs the hierachical generation of spins. It takes the cube with duplicated faces, finds the borders, and generates the cross. The cross is returned as a flat array of spins and is put into the cubes using the \texttt{make\_cross} function. Next, the cube is split into eight smaller cubes using the \texttt{divide\_into\_cubes} function, and the process is repeated until we obtain the cubes of size $3^3$. At this stage, the central spin in each cube is generated using the heatbath algorithm. At each level of the hierarchy, a different model from the provided list is used for the generation, and the arguments passed to the models (border and cross) are recorded in 
\texttt{list\_args\_for\_net}. The cubes are concatenated along the batch dimension, so the functions return \texttt{batch\_size}$\times8^d$ configurations of size $3^3$ where $d$ is the number of levels.}, label=lst:han]
def han(sample, nets, list_args_for_nets, device) -> tuple[torch.Tensor, list[list[torch.Tensor]]]:
    b_size = sample.shape[0]
    l = sample.shape[2] - 1
    for net in nets:
        l //= 2
        b_size *= 8
        border = find_borders(sample)
        cross = net.sample(border)
        list_args_for_nets.append([cross, border])
        sample += make_cross(
            cross, sample.shape[2], sample.shape[3], sample.shape[4], device)
        sample = torch.cat(divide_into_cubes(sample), dim=0)
    return sample, list_args_for_nets    
\end{lstlisting}

\begin{lstlisting}[float, caption={The function returns a tensor consisting of the boundary spins of given cube. The edges of the cube are discarded, as we are interested only in the spins that have nearest neighbours in the interior of the cube.}, label=lst:borders]
def find_borders(cube: torch.Tensor) -> torch.Tensor:
    assert (cube.shape[2]==cube.shape[3]==cube.shape[4])
    t = cube.shape[2] - 1
    borders = torch.cat(
        [cube[:, :, 0, 1:t, 1:t],
        cube[:, :, 1:t, t, 1:t],
        cube[:, :, t, 1:t, 1:t],
        cube[:, :, 1:t, 0, 1:t],
        cube[:, :, 1:t, 1:t, 0],
        cube[:, :, 1:t, 1:t, t]],
        dim=2)
    return borders
\end{lstlisting}

\begin{lstlisting}[float, caption={The function takes as input a flat tensor of spins without duplicates, constructs a 3D cross-like structure, and embeds it into an empty (filled with zeros) cube.}, label=lst:cross]
def make_cross(cross: torch.Tensor, L: int, device: torch.device):
    assert is_odd(L)
    cube = torch.zeros(
        [cross.shape[0], 1, L, L, L], device=device, dtype=torch.float32
    )
    arm_len = L - 2
    arm_len2 = (arm_len - 1) // 2

    xz_size = arm_len * arm_len;
    yz_half_size = arm_len2 * arm_len
    yx_quater_size = arm_len2 * arm_len2
    cross_parts = torch.split(
        cross,
        [
            xz_size,
            yz_half_size, yz_half_size,
            yx_quater_size, yx_quater_size, yx_quater_size, yx_quater_size
        ],
        dim=2,
    )

    # Loading XZ square (L, L)
    i = 0
    cube[:, :, 1: arm_len + 1, L // 2, 1: arm_len + 1] = cross_parts[i].reshape(
        cross.shape[0], 1, arm_len, arm_len)
    i += 1
    # Loading the two YZ rectangles (L, (L-1)/2)
    for y in range(2):
        offy = y * (arm_len2 + 1)
        cube[:, :, L // 2, 1 + offy: arm_len2 + 1 + offy, 1: arm_len + 1] = (
            cross_parts[i].reshape(cross.shape[0], 1, arm_len2, arm_len))
        i += 1
    # Loading the four XY squares ((L-1)/2, (L-1)/2)
    for x, y in product(range(2), range(2)):
        offx = x * (arm_len2 + 1)
        offy = y * (arm_len2 + 1)
        cube[:, :,
        offx + 1: offx + arm_len2 + 1, offy + 1: offy + arm_len2 + 1, L // 2] = (
            cross_parts[i].reshape(cross.shape[0], 1, arm_len2, arm_len2))
        i += 1

    return cube
  
\end{lstlisting}

\begin{lstlisting}[float, caption={The function divides a cube  into eights smaller cubes with the overlaping borders.}, label=lst:divide]
def divide_into_cubes(cube: torch.Tensor) -> list[torch.Tensor]:
    assert cube.shape[2] == cube.shape[3] == cube.shape[4]
    cube_size = cube.shape[2]
    small_cube_size = (cube_size + 1) // 2
    begin = (0, small_cube_size - 1)
    end = (small_cube_size, cube_size)

    return [cube[:, :, begin[x]:end[x], begin[y]:end[y], begin[z]:end[z]]
            for z in range(2) for x in range(2) for y in range(2)]
\end{lstlisting}

\begin{lstlisting}[float, caption = {The function
peforming the inverse of the \texttt{divide\_into\_cubes}
function: it takes eight cubes and puts them back into one large cube.}, label=lst:add_into_cube]
def add_into_cube(eight_cubes: Sequence[torch.Tensor]):
    small_cube_size = eight_cubes[0].shape[2]

    cube_size = 2 * small_cube_size - 1
    batch_size = eight_cubes[0].shape[0]
    cube = torch.zeros(batch_size, 1, cube_size, cube_size, cube_size, device=eight_cubes[0].device)

    begin = (0, small_cube_size - 1)
    end = (small_cube_size, cube_size)
    i = 0
    for z in range(2):
        for x in range(2):
            for y in range(2):
                cube[:, :, begin[x]:end[x], begin[y]:end[y], begin[z]:end[z]] = eight_cubes[i]
                i += 1

    return cube   
\end{lstlisting}

\bibliography{references2}

\end{document}